# Efficient Candidacy Reduction For Frequent Pattern Mining

M.H Nadimi-Shahraki[1], Norwati Mustapha[2], Md Nasir B Sulaiman[2], Ali B Mamat[2]

[1] Faculty of Computer Engineering, Islamic Azad University, Najafabad branch, Iran,
And Ph.D. Candidate of Computer Science, University of Putra Malaysia
[2] Faculty of Computer Science and Information Technology,
University of Putra Malaysia (UPM), Selangor, Malaysia.

*Abstract*— **Certainly, nowadays knowledge discovery or extracting knowledge from large amount of data is a desirable task in competitive businesses. Data mining is a main step in knowledge discovery process. Meanwhile frequent patterns play central role in data mining tasks such as clustering, classification, and association analysis. Identifying all frequent patterns is the most time consuming process due to a massive number of candidate patterns. For the past decade there have been an increasing number of efficient algorithms to mine the frequent patterns. However reducing the number of candidate patterns and comparisons for support counting are still two problems in this field which have made the frequent pattern mining one of the active research themes in data mining. A reasonable solution is identifying a small candidate pattern set from which can generate all frequent patterns. In this paper, a method is proposed based on a new candidate set called candidate head set or H which forms a small set of candidate patterns. The experimental results verify the accuracy of the proposed method and reduction of the number of candidate patterns and comparisons.**

*Keywords- Data mining; Frequent patterns; Maximal frequent patterns; Candidate pattern*

## I. INTRODUCTION

The explosive growth of data in all business, government and scientific applications creates enormous hidden knowledge in their databases. Certainly, in this decade knowledge discovery or extracting knowledge from large amount of data is a desirable task in competitive businesses. For example, daily a large amount of purchase data called market basket transactions are collected in the cashier counters of huge markets. The market management systems are interested in analyzing the purchase data to understand more about the behavior of their customers. The association analysis can represent interesting relationships hidden in large data set in the form of association rules. For example, 75% of customers who buy diapers also buy orange juice. These rules can be used to identify new opportunities for cross-selling markets' products to their customers. The association analysis is also useful in other applications such as web mining scientific applications. Data mining therefore appears to address the need of sifting useful information such as interesting relationships hidden in large databases. As shown in Figure 1, data mining is an essential step in the process of knowledge discovery from data (KDD) to extract data patterns. It is a composite process of multiple disciplines including statistics, database systems, machine learning, intelligent computing and information technology.

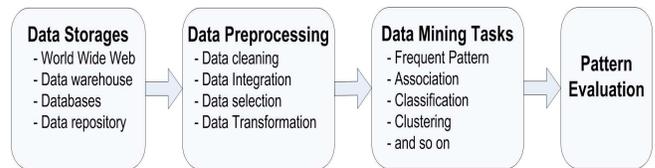

Figure 1. Data mining as a main step in KDD process

Since the first introducing [2], frequent patterns mining plays an important role in data mining tasks such as clustering, classification, prediction and especially association analysis. Frequent patterns are itemsets or substructures that exist in a data set with frequency no less than a user specified threshold. Identifying all frequent patterns is the most time consuming process due to a massive number of candidate patterns. In general, $2^i$-1 candidate patterns can be generated from a data set contains $i$ items. Therefore the computational requirements for frequent patterns mining are very expensive.

For the past decade there have been an increasing number of efficient algorithms to mine the frequent patterns by satisfying the minimum support threshold. They are almost based on three fundamental frequent patterns mining methodologies: Apriori, FP-tree and Eclat [9]. The Apriori-based algorithms significantly reduce the size of candidate sets using the Apriori principle that says all subsets of an infrequent itemset must be infrequent. But they still suffer from the generate-and-test strategy. They mine frequent patterns by generating candidates and checking their frequency against the transaction database. The FP-tree keeps only frequent items respect to the minimum support threshold by two database scan. Recently some FP-tree-based algorithms have been developed to capture the content of the transaction database only by one database scan which can be very useful for incremental updating of frequent patterns [12]. They usually traverse the tree to mine frequent patterns without candidate generation in same fashion. Only a few of them can fit the content of the transaction database in memory to eliminate the database rescanning and the mining model restructuring.

Meanwhile reducing the number of candidate patterns and the comparisons for support counting are still two problems in this field which have made the frequent pattern mining one of the active research themes in data mining. A reasonable solution is identifying a small candidate pattern set from which can generate all frequent patterns. In this paper, a method is proposed based on a new candidate set called candidate head set or H which form a small set of candidate patterns. It is an improvement of our previous method presented for maximal frequent pattern mining [13]. The proposed method is based on prime number characteristics for frequent pattern mining including a data transformation technique, an efficient tree structure called Prime-based encoded and Compressed Tree or PC_Tree and mining algorithm PC_Miner. The salient difference is that mining process makes use of the candidate head set and its properties to reduce the number of the candidate sets and comparisons. The PC_Miner algorithm strives to find long promising patterns during of the initial steps based on the candidate head set and its properties. Consequently, it prunes the search space efficiently.

The rest of this paper is organized as follows. Section 2 introduces the problem and reviews some efficient related works. The proposed method is described in section 3. The experimental results and evaluation show in section 4. Finally, section 5 contains the conclusions and future works.

## II. PROBLEM DEFINITION AND RELATED WORK

Simply, frequent patterns are itemsets or substructures that exist in a dataset with frequency no less than a user specified threshold. The first definition of frequent itemset was introduced for mining transaction databases (Agrawal et al. 1993).

### A. Problem Definition

Let L= $\{i_1, i_2 \ldots i_n\}$ be a set of items and D be a transaction database where each transaction T is a set of items and |D| be the number of transactions in D. Given X= $\{i_j \ldots i_k\}$ be a subset of L ($j \leq k$ and $1 \leq j, k \leq n$) is called a pattern. The support of the pattern X or Sup (X) in D is the number of transactions in D that contains X. The pattern X will be called frequent if its support is no less than a user specified support threshold min_sup σ ($0 \leq σ \leq |D|$). The problem of frequent pattern mining is finding all frequent patterns from dataset D with respect to specified min_sup σ. Various kinds of frequent patterns can be mined from different kinds of data sets. In this research, we use itemsets (sets of items) as a data set and the proposed method is for frequent itemset mining, that is, the mining of frequent itemsets from transactional data sets. However, it can be extended for other kinds of frequent patterns.

The complexity of frequent patterns mining from a large amount of data is generating a huge number of patterns satisfying the minimum support threshold, especially when min_sup σ is specified low. This is because, all sub-pattern of a frequent pattern are frequent as well. Therefore a long pattern contains a number of shorter frequent sub-patterns. Given the itemset lattice shown in Figure 2, which presents the list of all possible itemsets for L= {A, B, C, D, E}. A brute-force approach for mining frequent itemsets is to count the support every candidate itemsets in the lattice structure. It needs to compare each itemsets against every transaction. Obviously, this approach can be very expensive and it needs O (NML) comparisons, where N is the number of transactions, M is equal $2^i$-1 candidate itemsets for i items and L is maximum length of transactions. Therefore there are two main ideas to reduce the computational complexity of frequent itemsets mining. Firstly, reducing the number of candidate itemsets and secondly reducing the number of transactions those must be compared to count the support of the candidate itemsets.

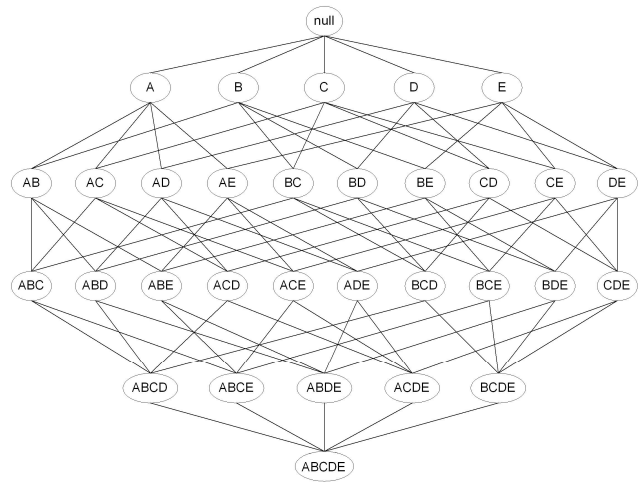

Figure 2. The itemset lattice for item set L= {A, B, C, D, E}

### B. Related work the

Agrawal and Srikant [3] introduced an interesting property called Apriori principle, to reduce the number of candidate sets among frequent k-itemsets: if a k-itemset is frequent then all of its sub-itemsets must also be frequent. The Apriori algorithm find the frequent 1-itemsets by first scanning the database , then generating the candidate frequent 2-itemsets by using the frequent 1-itemsets, and check them against the database to find the frequent 2-itemsets. This process is iterates until it can not generate any frequent k-itemsets for some k. The Apriori algorithm has been extended by several extension for improving efficiency and scalability.

The most important improvement of the Apriori which introduced new technique or method such as partitioning technique [16], hashing technique [14], sampling approach [17], dynamic itemset counting [5], use the Apriori principle approach as well. In many cases, the Apriori approach showed a good performance to reduce the size of candidate sets. However, in condition with a large number of frequent patterns or low minimum support thresholds, it almost suffers inherently from two problems; multiple database scans that are costly and generating lots of candidates [9].

Han et al. [8] proposed frequent pattern tree or FP-Tree as a prefix-based tree structure, and an algorithm called FP-growth. The FP-Tree stores only the frequent items in a frequency-descending order. The highly compact nature of FP-tree enhances the performance of the FP-growth. The FP-Tree construction requires two data scans. The FP-growth unlike the Apriori algorithm mines the complete set of frequent patterns without candidate generation. There have been introduced many extensions based on the FP-Tree approach such as depth-first generation by Agarwal et al. [1], H-Mine by Pei et al. [15], array-based implementation of prefix-tree-structure by Grahne and Zhu [7] and CanTree by leung et al [10]. The experimental results showed that FP-Tree and almost all its extensions have a high compactness rate for dense data set. However, they need a large amount of memory for sparse data set where probability for sharing common paths is low [10, 12].

The presentation of data which will be mined is an essential consideration in almost all algorithms. The mining algorithms can be classified according to two horizontal and vertical database layouts. Both the Apriori and FP-growth methods use horizontal data format (i.e., {TID: itemset}) to mine frequent patterns. Zaki [18] proposed Eclat algorithm or Equivalence CLASS Transformation by using the vertical data format (i.e., {item: TID_set}). The Eclat uses the lattice theory to represent the database items. The results showed that Eclat outperforms Apriori significantly. However, it needs an additional conversion step. This is because most databases use a horizontal format. Moreover, it uses a Boolean power set lattice that needs to much space to store the labels and tid-lists. Consequently, there have been introduced some efficient algorithms based on vertical layout. The Flex [11] is a lexicographic tree designed in vertical layout to store pattern X and list of transaction identifier where pattern X appears. Its structure is restricted test-and-generation instead of Apriori-like is restricted generation-and-test. Thus nodes generated are certainly frequent. The Flex tree is constructed in depth-first fashion. The experimental results showed the Flex is an efficient algorithm to find long and maximal frequent patterns. However, it needs a large amount of memory especially to store the list of transaction identifier.

## III. PROPOSED METHOD

Based on related works' results, reducing the number of candidate sets and comparisons (to count the support) are two effective way to enhance the performance of mining process. They showed using the Apriori principle can reduce the number of candidate sets and well-organized tree structure such as FP-Tree which captures the content of the transaction database reduces the number of comparisons of the support counting. Therefore, in this research, we aim to use both the Apriori principle and well-organized tree structure in our proposed method. We proposed a method by using a simple and effective tree structure for maximal frequent pattern mining which can capture all content of the transaction database [13]. This research proposes an improvement of previous version using a new efficient candidate set called candidate head set based on the Apriori principle to reduce the number of candidate sets. The proposed method is also based on prime number characteristics including a data transformation technique, an efficient tree structure called Prime-based encoded and Compressed Tree or PC_Tree and mining algorithm PC_Miner. The salient difference is that mining process makes use of the candidate head set and its properties to reduce the number of candidate sets and comparisons. In fact the PC_Miner algorithm prune the search space by using the candidate head set and it finds the most promising candidate set efficiently. This section is followed by reviewing of the data transformation technique and the PC_Tree. Then the candidate head set and its properties are explained to show how they can use in the PC_Miner algorithm to reduce the number of candidate sets.

### A. Data Transformation Technique

As shown in Fig 1 the data transformation is an essential process in data preprocessing step which can reduce the size of database. Obviously, reducing of the size of database can enhance the performance of mining algorithms. Our method uses a prime-based data transformation technique to reduce the size of transaction database. It transforms each transaction into a positive integer called Transaction Value (TV) during of the PC_Tree construction as follows: Given transaction T = (tid, X) where tid is the transaction-id and X = $\{i_j \ldots i_k\}$ is the transaction-items or pattern X. While the PC_Tree algorithm scans transaction T, the transformer procedure considers a prime number $p_r$ for each item $i_r$ in pattern X, and then $TV_{tid}$ is computed by Equation 1 where T= (tid, X), X = $\{i_j \ldots i_k\}$ and $i_r$ is presented by $p_r$.

$$TV_{tid} = \prod_{j}^{k} p_r \qquad (1)$$

The data transformation technique utilizes Equation 1 based on simple following definitions:

"A positive integer N can be expressed by unique product N = $p_1^{m_1} \, p_2^{m_2} \ldots p_r^{m_r}$ where $p_i$ is prime number, $p_1 \prec p_2 \prec \cdots p_r$ and $m_i$ is a positive integer, called the multiplicity of $p_i$" [6].

For example, N = 1800=$2^3*3^2*5^2$. Fundamentally, there is no duplicated item in transaction T. Hence we restrict the multiplicity only to $m_i = 1$ without losing any significant information. Therefore N can be produced by $p_1 p_2 \cdots p_r$. To facilitate the transformation process used in our method, let's examine it through an example. Let item set L= {A, B, C, D, E, F} and the transaction database, DB, be the first two columns of Table 1 with eight transactions. The fourth column of Table 1 shows $TV_{tid}$ transformed for all transactions. Generally the average length of items used in the benchmark datasets is smaller than those in the real applications. For example customer purchase transaction $T_C$= (3, {55123450, 55123452, 55123458}) from a market can be presented by third transaction T= (*3, {A, B, E}*) in Table 1. Although, the length of items in transaction $T_C$ is bigger than T but both T and $T_C$ can be transformed into the same TV 66 by using our data transformation technique. Hence it is an item-length independent transformation technique. The experiments showed that by applying this data transformation technique, the size of real transaction databases can be reduced more than half [13].

| TID | Items | Transformed | TV |
|-----|-------|-------------|-----|
| 1 | A, B, C, D, E | 2, 3, 5, 7, 11 | 2310 |
| 2 | A, B, C, D, F | 2, 3, 5, 7, 13 | 2730 |
| 3 | A, B, E | 2, 3, 11 | 66 |
| 4 | A, C, D, E | 2, 5, 7, 11 | 770 |
| 5 | C, D, F | 5, 7, 13 | 455 |
| 6 | A, C, D, F | 2, 5, 7, 13 | 910 |
| 7 | A, C, D | 2, 5, 7 | 70 |
| 8 | C, D, F | 5, 7, 13 | 455 |

### B. PC_Tree construction

There have been introduced several methods to reduce the complexity of frequent pattern mining process using well-organized tree structure. Thus the tree structures have been considered as a basic structure in previous data mining research [8, 10-13]. Recently, we introduced a novel tree structure called Prime-based encoded and Compressed Tree or PC_Tree [13].It is very simple but still powerful to capture the content of transaction database efficiently. Unlike the previous methods, the PC_Tree is based on prime number characteristics. Moreover PC_Tree has some nice properties which used to prune the search space during of mining process. Let's review the PC_Tree.

A PC_Tree consists of one root labeled as "null" and some nodes that form sub trees as children of the root. The node structure consists of several fields: value, local-count, global-count, status and link. The value field stores the TV made by the data transformation technique during of insertion procedure in the PC_Tree construction algorithm. In fact, the value field registers which transaction this node represents. The local-count field registers the number of individual transaction represented by its node in the whole of transaction database. It is set to 1 during inserting current TV in a new node and or if there is a node with same TV then the value of its local-count field is increased by 1. Hence there is no duplicated TV in the PC_Tree. The global-count field registers frequency of its TV in its sub tree (descendant) to use in the support computing function of its TV. The status field is to keep tracking of traversing which is changed from 0 to 1 when a node visited in the traversing procedure. The link field is to form sub trees.

The PC_Tree construction algorithm forms a PC_Tree by inserting TV(s) based on definitions below:

***Definition 1:*** TV of the root is assumed null and can be divided by all TVs.

***Definition 2:*** Sub tree $R_i = (root, n_0, n_1, ..., n_i)$ is a sub descendant if only if

$TV(n_j) | R_j = (root, n_0, n_1, ..., n_j)$ *where* $0 \le j \le i$. It means any TV stored in nodes of the sub descendant $R_j = (root, n_0, n_1, ..., n_j)$ can be divided by TV stored in node $n_j$.

***Definition 3:*** Sub descendant $R_i = (root, n_0, n_1, ..., n_i)$ is a descendant if $n_i$ is a leaf.

***Definition 4:*** In a PC_Tree, If TV $(n_r)$ = TV $(n_s)$ then $r = s$. The insertion procedure increases local-count field of node $n_r$ by 1 if the current TV is equal with TV of $n_r$.

Based on definition 1-4, the PC_Tree can be constructed by algorithm 1 as follow:

**Algorithm 1** (PC_Tree construction).
**Input**: A transaction database *DB*.
**Output**: The PC_Tree of *DB*.
**Method**: The PC_Tree is constructed as follows.

1. Create the root of the PC_Tree and label it as "null".
2. For each transaction *T* in *DB*, insert *T* into the PC_Tree as follow:
   2.1. Scan *T* from input file, transform *T* into its *TV* and update the item frequency table.
   2.2. Add *TV* in the tree as follows.
      2.2.1. If *TV* can be an element of existent descendant *R* then insert *TV* as follows. If there is a node in *R* with same TV then increase its local-count and global-count fields and its children's global-count field by 1; else create a new node, with its local-count field initialized to 1. Link the new node to its parent and children. Increase its children's global-count field by 1 and set its global-count field by summation value of its local-count field and values of its parent's global-count field.
      2.2.2. Else; the *TV* cannot be an element of existent descendants. Create a new node, with its local-count and global-count initialized to 1. Link the new node to the root as its parent and to its children. Increase its children's global-count field by 1.

Figure 3 shows the PC_Tree constructed for transaction database shown in Table 1. Each node presents a TV or pattern followed by two numbers after ":" to indicate the local-count and global-count respectively. There are several important properties of PC_Tree that can be derived from the PC_Tree construction procedure.

**Property 1**: In the PC_Tree, all nodes are arranged according to TV-descending order.

**Property 2:** Important procedures used in the PC_Tree algorithm are almost done only by two simple mathematic operations *product and division*. Obviously using mathematic operations will enhance the performance instead of string operations.

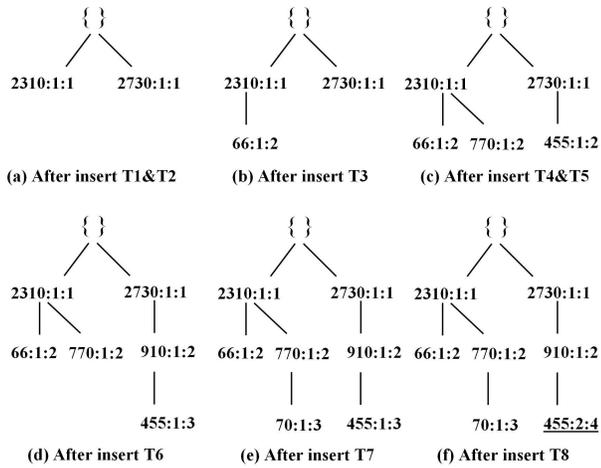

Figure 3.    Step by step PC_Tree construction

$H = \{ h_r \mid h_r$ is candidate head of $R_r$, $0 \le r \le j$ such that if $h_r$, $h_s \in H$ and $h_r \subseteq h_s \Rightarrow h_r = h_s \}$ is candidate head set of the PC_Tree.

Based on above definitions, the PC_Tree has some interesting properties which will facilitate frequent-pattern mining. Given $\sigma$ as min_sup and pattern P and Q have been represented by TV (P) and TV (Q) in descendant R respectively.

**Property 3:** $P, Q \in R$, $\sup(P) \prec \sup(Q)$ if and only if $TV(Q) \mid TV(P)$ (i.e. TV (P) can be divided by TV (Q)).

**Property 4:** sup (P) $\ge \sigma$ and sup (P's parents in its descendants) $\prec \sigma$ if and only if P is a maximal frequent pattern.

Now, this is possible to use the candidate head set to find a smaller candidate set from which the maximal frequent patterns can be derived. Consequently all frequent patterns can be generated by the maximal frequent pattern set. Therefore, based on the candidate head set definition and above properties, we have the following algorithm for frequent pattern mining using PC_Tree.

---

*C.    PC_Miner algorithm using candidate head set*

In our previous work [13], the PC_Tree was mined by the PC_Miner algorithm in a top-down traversing fashion. It finds the maximal frequent patterns as the smallest representative set from which all frequent patterns can be derived. The PC_Miner makes use from superset and subset pruning to enhance the performance of maximal frequent pattern mining. The weakness of PC_Miner is that it considers all sub trees or descendants with same possibility of traversing. To solve this weakness, the PC_Miner is improved by an efficient heuristic called candidate head set denote H to reduce the effective branch factor of the PC_Tree as follows.

Without considering the root, let $R_r = (n_0, n_1, ..., n_i)$ be a descendant in the PC_Tree, the node $n_0$ as the first node of the descendant $R_r$ that registers the biggest TV in $R_r$ is called head of the descendant $R_r$. In other word the immediate children of the root make the head set.

***Definition 6 – candidate head:*** the positive integer $h = p_1 \ldots p_j$ which made by product prime numbers $p_1 \ldots p_j$ is the candidate head of the descendant $R_r$ if only if $h$ is the largest subset of $n_0$ and $\sup(p_k) \ge$ min_sup, $1 \le k \le j$.

According to the data transformation technique used in the PC_Tree construction, TV $(n_0)$ is a product of prime numbers $p_1 \ldots p_k$ which their frequency register in the item frequency table(see Table 2) by step 2.1 of the algorithm1. Using this table and the minimum support threshold, infrequent prime numbers can be removed from TV $(n_0)$ as head of descendant $R_r$ to make its candidate head.

***Definition 7 – candidate head set:*** Let $R = (R_0, ..., R_j)$ be the descendant set of the PC_Tree,

---

**Algorithm 2** (PC_Miner: Mining frequent patterns from the PC_Tree by candidate head set)
**Input**: A transaction database DB, represented by PC_Tree and minimum support threshold $\sigma$.
**Output**: The complete set of frequent Patterns of DB.
**Method**: The PC_Miner mines frequent patterns as follows.
1:  Make the candidate head set H by using the item frequency table.
2:  Let $k_{max}$ denote the maximum size of itemsets in H.
3:  $F = F_1 = \{ f \mid \sup(f) \ge$ min_sup, $\mid f \models 1 \}$
4:  $k = k_{max}$
5:  for $k$ downto 2 do
6:   $H_k = \{ f \mid f \in H, \mid f \models k \}$.
7:    for each $f \in H_k$ and $f \notin F$ do
8:     if Sup $(f) \ge$ min_sup  *// property 4 and corollary 1 //*
9:       F=F U all subsets of $f$.
10:    else if k>2 then
11:       add all (k-1)-subsets of $f$ to H.
12:    end if
13:   end for
14: end for

The correctness and completeness of the process in the PC_Miner algorithm should be justified. This is accomplished by first introducing important lemma and corollary as follow:

**Lemma 1:** Let $H = \bigcup\limits_{i=1}^{k} H_i$ consist of all k-candidate head sets, then the complete set of the frequent itemsets can be generated by $H$.

**Rationale:** Let F be the frequent items in DB, n be head of the descendant R and h (n) is the set of frequent items in n, i.e., h (n) = n ∩ F. According to candidate head set definition, h (n) is the candidate head of the descendant R and belongs to $H$. In other word the candidate head set $H$ used in the PC_Miner algorithm includes all frequent items. The relationship among the candidate head set H, frequent and maximal frequent itemsets are shown in Figure 4.

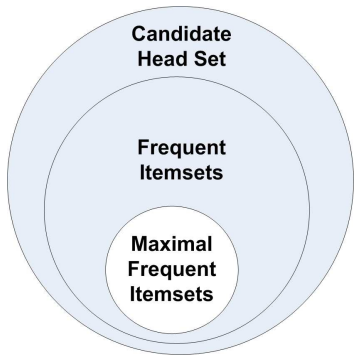

Figure 4.   Relationship among candidate head set, frequent and maximal frequent itemsets

**Corollary 1:** The complete set of the frequent itemsets F can be generated by $(\bigcup\limits_{i=2}^{k} H_i) \bigcup F_1$ where $F_1$ is set of 1-length frequent itemsets which can be derived from item frequency table made in the PC_Tree algorithm.

The number of the candidate head sets is very smaller than the frequent and even maximal frequent itemsets. However, Figure 4 shows that the candidate head set is a superset of frequent and maximal itemsets. To illustrate let's examine the mining process. In the PC_Miner algorithm, the line 3 makes 1-frequent itemsets by using the item frequency table which made during the construction process. In outer for-loop between lines 5-14, all k-candidate heads (k>2) are made and investigated respectively. Then during of line 7-13 all frequent itemsets are generated by using maximal frequent itemsets found in line 8 based on property 4. According to the relationship shown in Figure 4 and steps of the PC_Miner algorithm which investigate all candidate heads; therefore there is no missing frequent itemsets.

The PC_Miner algorithm make use from superset infrequent pruning based on the candidate head set introduced in this research which is an effective way to eliminate some of the candidate itemsets. In fact the candidate head set definition is based on the Apriori principle which all subsets of a frequent itemset must also be frequent. For instance there are $2^6 - 1 = 63$ possible candidate itemsets in the transaction database DB presented in Table1. The PC_Miner uses the frequency of items computed in step 2.1 of the PC_Tree algorithm to find the candidate head set. Table 2 shows the frequency of items of DB. Consequently, items B and E are infrequent when min_sup=4. Therefore, the PC_Miner starts mining with candidate head set H includes only 4-candidate itemset ACDF and the PC_Miner only examined 6 candidate itemsets {ACDF, ACD, ACF, ADF, CDF, AF}. Above definitions and PC_Miner algorithm show that using the candidate head set can reduce the number of candidate itemsets. Moreover, the experimental results support the accuracy and efficiency of our method.

## IV.   EXPERIMENTAL RESULTS

In this section, we evaluate the accuracy and performance of our method by several experiments. All experiments were performed in a time-sharing environment in a 2.4 GHz PC with 2 GB memory. We used several sparse and dense datasets which used in previous works as benchmark datasets. The synthetic sparse datasets are generated by the program developed at IBM Almaden Research Center [3] and real dense datasets are download from UC Irvine Machine Learning Repository [4]. The results reported in figures were computed by the average of multiple runs. According to the space limitation and the problem specifications, only the results of experiments by using synthetic sparse dataset T10I5D100K and real dense dataset mushroom which are the most popular benchmark datasets in this field are presented in this paper. The number of transactions, the average transaction length, the number of items and the average frequent pattern length of T10I5D100k are set to 100k, 10, 1000 and 4 respectively. The mushroom dataset consists of the characteristics of various mushroom species. The number of records, the number of items and the average record length are set to 8124, 119 and 23 respectively.

The first experiment is to proof the accuracy of the proposed method. However the correctness and completeness of the process in the PC_Miner algorithm was justified in previous section. The number of frequent patterns mined by the PC_Miner algorithm versus support using T10I5D100k and mushroom datasets is shown in Figures 5 and 6 respectively. We compared all frequent patterns mined by the proposed method using the candidate head set with those mined by the Apriori algorithm in several datasets. They were exactly equal.

The second experiment evaluates the number of the candidate sets generated by the PC_Miner and Apriori which is one of the most efficient algorithms in term of candidacy reduction. Figure 7 shows the number of candidate sets generated by PC_Miner and Apriori algorithms versus different minimum support thresholds over dataset T10I5D100k.

The efficiency of using the candidate head set in the PC_Miner algorithm is verified by third experiment which compares the run time versus the support. Figure 8 and 9 present the efficiency of the PC_Miner algorithm using the candidate head set versus the Apriori algorithm in T10I5D100k and mushroom datasets over different minimum support thresholds respectively.

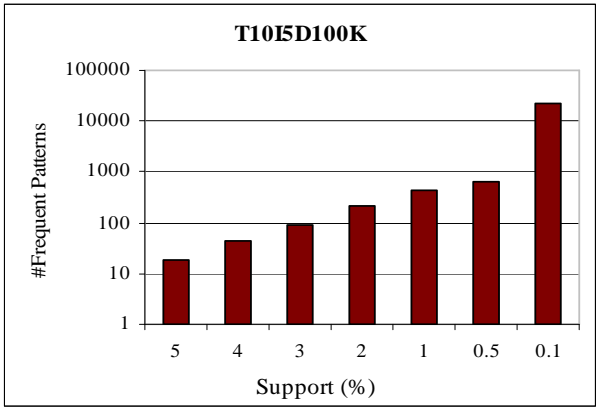

Figure 5.   #Frequent Patterns Vs. Support

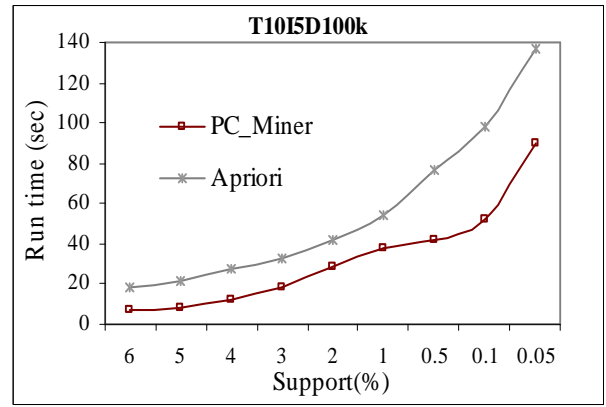

Figure 8.   Run time Vs. Support

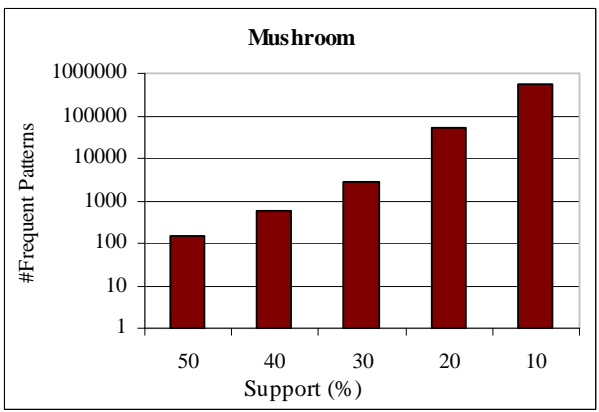

Figure 6.   #Frequent Patterns Vs. Support

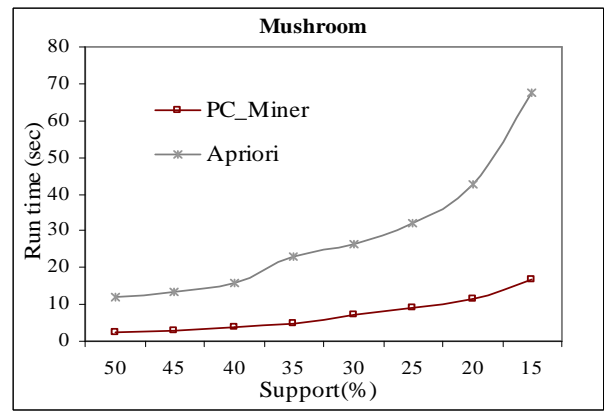

Figure 9.   Run time Vs. Support

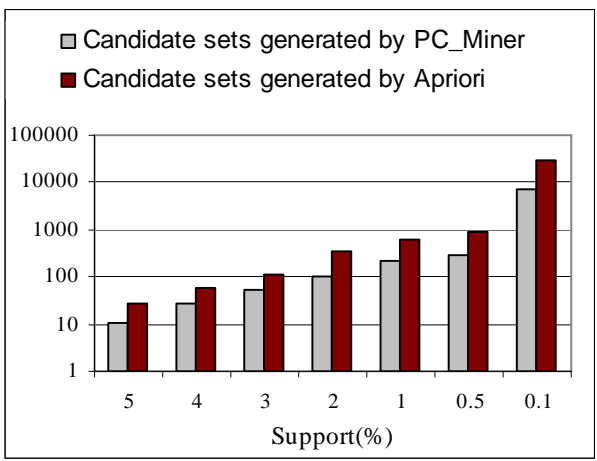

Figure 7.   The number of  candidate sets Vs. Support

## V.   CONCLUSION AND FUTURE WORKS

In this paper we introduced the candidate head set to reduce the number of candidate sets in mining process. Our previous method [13] was improved by using the candidate head sets to propose an efficient method for frequent pattern mining. The experimental results verified the accuracy and efficiency comparing with the Apriori algorithm which is one of the most efficient algorithms in term of candidacy reduction.

Particularly, we introduced a new method based on prime number characteristics using candidate head sets to find completed frequent patterns by using maximal frequent patterns. The proposed method can be improved for incremental mining of frequent patterns where database transactions can be inserted, deleted, and/or modified incrementally. Moreover it can be improved for interactive mining of frequent patterns where minimum support threshold can be changed to find new correlation between patterns without rerunning the mining process from scratch.